\documentstyle[prl,aps,graphics,epsfig]{revtex}

\input{psfig.sty}
\begin{document}
\pagestyle{empty}
\draft
\twocolumn
\wideabs{
\title{New, Efficient and Robust, Fiber-Based Quantum Key Distribution Schemes}
\author{W.T. Buttler, J. R. Torgerson and S. K. Lamoreaux}
\address{University of California, Los Alamos National Laboratory, Los Alamos, 
New Mexico 87545}
\date{\today}
\maketitle
\begin{abstract}
We present a new fiber based quantum key distribution (QKD) scheme which can be 
regarded as a modification of an idea proposed by Inoue, Waks and Yamamoto (IWY) 
\cite{ref:iwy01}. The scheme described here uses a single phase modulator and 
two differential delay elements in series at the transmitter that form an 
interferometer when combined with a third differential delay element at the 
receiver. The protocol is characterized by a high efficiency, reduced exposure 
to an attack by an eavesdropper, and higher sensitivity to such an attack when 
compared to other QKD schemes. For example, the efficiency with which 
transmitted data contribute to the private key is $3/4$ compared with $1/4$ for 
BB84 \cite{ref:bb84}. Moreover, an eavesdropper can aquire a maximum of $1/3$ of 
the key which leads to an error probability in the private key of $1/3$. This 
can be compared to $1/2$ and $1/4$ for these same parameters in both BB84 and 
IWY. The combination of these considerations should lead to increased range and 
key distribution rate over present fiber-based QKD schemes.
\end{abstract}
\pacs{PACS Numbers: 03.67.-a, 03.67.Hk, 03.67.Dd, 42.79.Sz}
}
\narrowtext

The quantum key distribution (QKD) scheme recently proposed by Inoue, Waks and 
Yamamoto \cite{ref:iwy01} represents a new concept that should prove to be a 
significant advance for fiber-based QKD. Their scheme, which we label IWY, is 
sketched in Fig.~\ref{fig:iwy01} for the case of a two delay-element 
transmitter. Essentially, a single pulse entering the transmitter leaves as a 
superposition of three pulses,  the last two of which have been given a random 
phase shift of $\Delta \phi \in \{0, \pi\}$ relative to the first in addition to 
a fixed time delay of $\Delta t$ and $2 \Delta t$ relative the time associated 
with the input state: The photon pulse is superposed in time and phase. The 
receiver effects a time delay of $\Delta t$ and causes the superposed pulses to 
overlap and hence interfere in the same manner as superposed pulses in the 
standard Franson interferometer \cite{ref:franson91} used in BB84-like fiber QKD 
as shown in Fig.~\ref{fig:bb84}. (The transmitter and receiver are typically 
referred to as Alice ({\bf A}) and Bob ({\bf B}) within the quantum information 
community.) 

The advantages of this new scheme are numerous: ({\rm I}) the efficiency with 
which transmitted data contribute to the private key, which we label $\eta_p$ 
the ``protocol efficiency,'' is increased to $\eta_p = 1 - 1/N$, where $N$ is 
the number of temporal superpositions leaving the transmitter; ({\rm II) the 
amount of information an eavesdropper ({\bf E}, usually Eve) can acquire on the 
key is reduced as $N$ increases; ({\rm III) the maximum error probability caused 
by eavesdropping, which we name ``disturbance'' and label as $p_d$, increases 
with $N$ and allows error reconciliation to be performed on data with a higher 
initial error probability; and ({\rm IV) {\bf B} does not need a phase-modulator 
(PM).

The implications of these advantages, relating to the points raised above, are 
that {\bf A} and {\bf B} generate more key and can tolerate and correct for 
higher error rates because: ({\rm i}) qubits accumulate at a higher rate; ({\rm 
ii}) the fraction of key leaked during eavesdropping, $\eta_e$, that must be 
discarded to ensure the privacy is reduced with increasing $N$ ($\eta_e$ relates 
the maximum amount of information {\bf E} can acquire on the quantum key in a 
specified attack on the quantum channel); ({\rm iii}) the liklihood of detecting 
{\bf E}'s attacks on the quantum channel is increased as her eavesdropping 
creates a larger disturbance within the qubits; and ({\rm iv}) the random choice 
of $\Delta \phi$ by {\bf B} necessary in other schemes is eliminated in IWY and 
yields an increase in $\eta_p$ over these other schemes ($\eta_p$ increases more 
than $300$\% for the case of the typical, or current, implementations of fiber 
based BB84, e.g. as shown in Fig.~\ref{fig:bb84}).

A further consequence of removing the PM from the receiver is that the typical 
$3$-$6$\,dB loss in PMs is removed from {\bf B}'s receiver. Thus, the key rate, 
robustness against errors, and distance over which qubits can be transmitted can 
be enhanced over currently realized fiber based QKD systems.

The scheme we introduce here, and labelled BLT and shown in Fig.~\ref{fig:blt01} 
for two transmitter delay elements, improves upon IWY in several significant 
ways while retaining all of IWY's advantages over other schemes. First, BLT is 
simpler to implement because the series-element construction of BLT can be 
achieved with simple $50/50$ beamsplitters (BSs), while the parallel 
construction of IWY requires more specialized BSs to balance the output pulses' 
intensities. Second, the efficiency ($\eta_p$), potential eavesdropper knowledge 
($\eta_e$) and disturbance ($p_d$) all favor BLT over IWY for the same number of 
delay elements in the transmitter. In BLT, the number $N$ of superposed pulses 
increases as $N = 2^m$ where $m$ is the number of delay elements at the 
transmitter; in IWY $N$ increases as $N = m + 1$. For example, the serial scheme 
of BLT yields $\eta_{p+} \ge 3/4$ while for the parallel scheme of IWY 
$\eta_{p\parallel} \ge 2/3$ with the lower bound given by $m = 2$ in each case. 
Moreover, the eavedropper's potential information $\eta_e$ is less in BLT than 
in IWY, and the disturbance $p_d$ is greater in BLT than in IWY, regardless of 
the type of attack employed by {\bf E}. Third, a single PM outside the 
transmitter delay elements further reduces the complexity and expense of the 
system and can increase its efficiency. For example, it is well-known that QKD 
is more secure and more efficient if each superposition of pulses contains 
exactly one photon. This would most likely be realized with a single photon 
source \cite{ref:photon-source} before the delay elements and appropriately 
modified optics as discussed below. A single PM outside the transmitter delay 
elements dramatically increases the receiver efficiency in this situation, again 
due to the typical $3$-$6$\,dB loss of PMs.

Consider a fiber BB84 system as shown in Fig.~\ref{fig:bb84} in which {\bf A} 
sends single, or weak coherent, photon states to  {\bf B} 
\cite{ref:franson91,ref:hughes48,ref:mhhtzg97}. For each input pulse the phase 
of {\bf A}'s PM is randomly toggled among a choice of four phases: $\phi_A \in 
\{0, \pi/2, \pi, 3 \pi/2 \}$. As the pulses arrive at {\bf B}, the PM at that 
end is toggled randomly between $\phi_B \in \{0, \pi/2 \}$ for each 
superposition of pulses to select the measurement basis for that superposition. 
({\bf B}'s delay element is adjusted so that a phase difference $\phi_A - \phi_B 
= 0$ causes always a photodetection at the same detector and $\phi_A - \phi_B = 
\pi$ causes always a photodetection at the other detector.) After the weak 
pulses are exchanged {\bf B} reveals the times when photodetections were 
observed and the measurement bases chosen at those times. {\bf A} notes when 
their phases satisfied $\phi_A - \phi_B \in \{0, \pi\}$, and tells {\bf B} to 
discard measurement results which were not collected when this condition was 
satified. What remains is potentially secret key in which $\phi_A - \phi_B \in 
\{0, \pi \}$ correspond to binary values $\{0, 1\}$.

The exchanged qubits will undoubtedly contain errors which must be eliminated 
through an appropriate error reconciliation protocol 
\cite{ref:bbbss92,ref:bs94}. After error reconciliation, privacy amplification 
\cite{ref:bbcm95} --- which is linked to the error rate --- must be applied to 
the remaining error-free key to reduce {\bf E}'s information to an acceptable 
level. If the error rate is too high all of the remaining key bits may be 
eliminated during this phase. In addition, the channel must be authenticated 
\cite{ref:sig-auth}; authentication represents an additional protocol expense.

It is worthwhile to note the fraction of key sent by {\bf A} which remains after
error reconciliation and privacy amplification for each of the schemes discussed 
here. We label this fraction $R_k$~\cite{ref:note4} and find
\begin{equation}
   R_k = \eta_p(\mu_r - \eta_e \frac{p_o}{p_d}) \text,
\end{equation}
where $p_o$ is the measured error probability per bit in the key (before error
reconciliation), and $\mu_r$ is the fraction of {\bf B}'s key material that
remains after error reconciliation, and depends on $p_o$. The fraction of
the distributed key that must be discarded through privacy maintenance is given 
by $\eta_e{p_o}/{p_d}$. Although an additional amount of key must be discarded 
if weak coherent states of light are used, this issue is disregarded in the 
discussion because $\bar n$ can be made arbitrarily small (in theory at least), 
and because other states of light can be employed that are not susceptible to 
this type of attack (e.g, entangled photon sources \cite{ref:E91,ref:kwiat2000}, 
or other single-photon states \cite{ref:photon-source}).

Consider the (typical) fiber based BB84 implementation described above, and 
assume that the PM transmission is unity. This implementation has a receiver 
efficiency $\eta_{p(bb84)} = 1/4$ \cite{ref:ideal-bb84}: $1/2$ from the 
$50$:$50$ BS at the input of {\bf B}'s delay element and $1/2$ due to the random 
bases choice that is crucial to BB84 security. The $\eta_e$ and $p_d$ depend on 
the type of attack used by {\bf E}. The most powerful, and yet realistic, attack 
that has been devised for this scheme is for {\bf E} to intercept and resend the 
photon pulses in the Breidbart Basis \cite{ref:bbbss92}. With this type of 
attack, $\eta_e \le 0.585$ and $p_d = 1/4$. If weak coherent pulses were used to 
transmit the key, another fraction $\bar{n}/4$ of the key is also potentially 
compromised to {\bf E}, where $\bar{n} \ll 1$ is the average number of photons 
per superposition.

The IWY and BLT schemes accomplish key exchange in a similar manner. That is, 
superposed photon pulses (single-photon or weak-coherent states) are transmitted 
from {\bf A} to {\bf B}. {\bf B}'s delay element is adjusted so that a phase 
difference $\Delta\phi\in\{0,\pi\}$ between adjacent pulses in a superposition 
always cause a photodetection at a particular detector. After photodetections 
are observed, {\bf B} tells {\bf A} at what times they were observed relative to 
an accepted time standard. Because {\bf A} knows what the phases of each 
superposition were, {\bf A} knows which of {\bf B}'s detectors registered a 
photodetection.

As mentioned previously, $\eta_{p\parallel}=2/3$ while $\eta_{p+}=3/4$ for the
implementations of BLT and IWY sketched in Figs.~\ref{fig:iwy01} and 
\ref{fig:blt01} respectively. The strongest attack that can be employed by
{\bf E} is likely an intercept/resend attack. However, unlike BB84, a 
Breidbart-like attack is not possible in IWY or BLT as the measurements are made 
in orthogonal bases. For these schemes, the strongest intercept/resend strategy 
is to build a receiver much like {\bf B}, but to replace the first BS with a 
fast switch (we do not specify the switch technology, but merely state it is 
possible). In this manner, adjacent pulses can be made to overlap without loss 
of signal due to the first or last pulse in the superposition contributing less 
to the interference than the central pulses. If exactly one photon exists in the 
superposition {\bf E} will measure one superposition phase difference 
($\Delta\phi$), but have no knowledge of the other remaining phase 
difference(s). Thus {\bf E}'s knowledge of the key for the superpositions that 
are intercepted and resent, and hence $\eta_e$ can be calculated as 
$\eta_{e\parallel} = 1/2$ and $\eta_{e+} = 1/3$ for the IWY and BLT 
respectively. The corresponding disturbances are $p_{d\parallel} = 1/4$ and 
$p_{d+} = 1/3$. If the effect of multiple photons in a superposition from the 
use of weak coherent states is included, {\bf E} can know another fraction of 
the key equal to $\bar n/4$ or $\bar n/6$ for IWY or BLT respectively.

An important variation to either the IWY or BLT scheme is to add a BS and an
extra delay element at the receiver as shown in (Fig.~\ref{fig:blt01v3}). For 
this system (denoted with $\oplus$), $\eta_{p\oplus} = 5/8$ while 
$\eta_{e\oplus} = 1/5$ is significantly reduced, and {\bf E}'s disturbance to 
the final key is increased to as much as $p_{d\oplus} = 2/5$, if the additional 
BS is equally reflecting and transmitting.

With these considerations, $\eta_p \in \{1/4, 2/3, 3/4,5/8\}$ and $\eta_e/{p_d} 
\in \{2.34, 2, 1, 1/2\}$ for \{BB84, IWY, BLT and BLT$\oplus$\} respectively. 
The rate at which key is generated is proportional to $R_k$ and can be 
approximated from  Fig.~\ref{fig:key-rate} where $R_k$ is plotted for the case 
of $\mu_r = 1 + (1 - p_o)\log_2(1 - p_o) + p_o\log_2(p_o)$: $1$ minus the {\it 
Shannon entropy}.

In any case, in a practical QKD system, $R_k$ will be bounded by the values 
plotted in Fig.~\ref{fig:key-rate} for the four schemes at the same $p_o$. 
Figure~\ref{fig:key-rate} highlights the value of the relative schemes presented 
here. For $p_o \lesssim 0.13$, it is clear that BLT has a great advantage over 
the other three schemes: qubits are collected at a higher rate than for the 
other $3$ schemes. For $p_o \gtrsim 0.13$, BLT$\oplus$ retains a significant 
advantage due to its reduced leakage of $\eta_e = 1/5$, and disturbance $p_d = 
2/5$. These facts do not consider the added complexity of BLT$\oplus$ at {\bf B} 
which includes the second delay element and additional detectors. For 
simplicity, efficiency, and protection against an intercept/resend attack by 
{\bf E}, BLT offers great advantages over BB84, IWY and BLT$\oplus$.

In conclusion, we have introduced two new quantum key distribution schemes for 
optical fiber systems based on the recently presented idea of Inoue, Waks and 
Yamamoto \cite{ref:iwy01}. Our new schemes are more efficient and robust against 
eavesdropping than other schemes proposed to date (BLT$\oplus$ is more efficient 
than IWY for error rates $\gtrsim 2$-$3$\%). In particular, BLT also has the 
advantage of being relatively easy and inexpensive to implement.

%
%
\begin{figure}[!h]
\psfig{figure=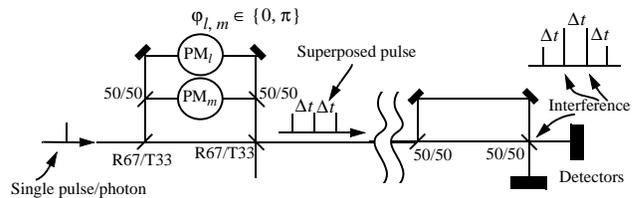,width=3.25 in}
\caption{The IWY parallel-delay element interferometer. Alice ({\bf A}) is on 
left, Bob ({\bf B}) on right. {\bf A} controls all phase information 
($\text{PM}_l$, and $\text{PM}_m$),  where the $l$ represents the ``longer'' 
path and $m$ represents the ``medium''  path (relative to the ``short,'' or 
direct path). {\bf A} randomly toggles $\phi_m$ and $\phi_l$ between values of 
$0$ and $\pi$.}
\label{fig:iwy01}
\end{figure}

\begin{figure}[!h]
\psfig{figure=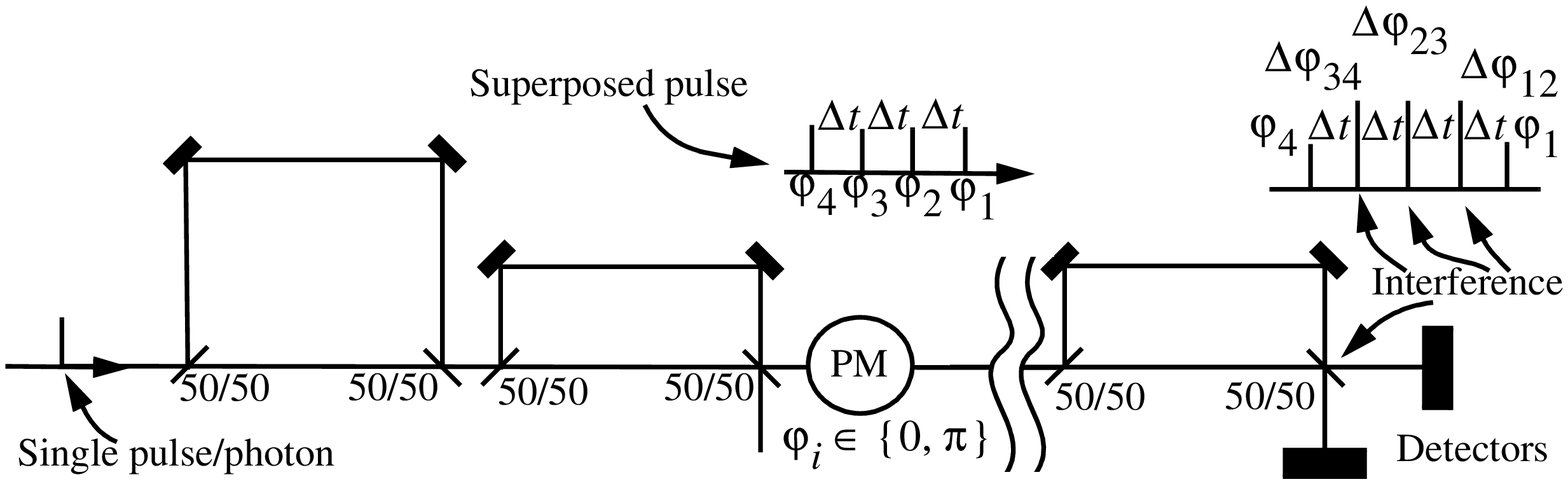,width=3.25 in}
\caption{The BLT series-delay element interferometer. {\bf A} is on left, {\bf 
B} on right. {\bf A}  controls all phase information by randomly adding a phase 
value of $0$ or $\pi$ to each of the four temporal superposition locations.}
\label{fig:blt01}
\end{figure}

\begin{figure}[!h]
\psfig{figure=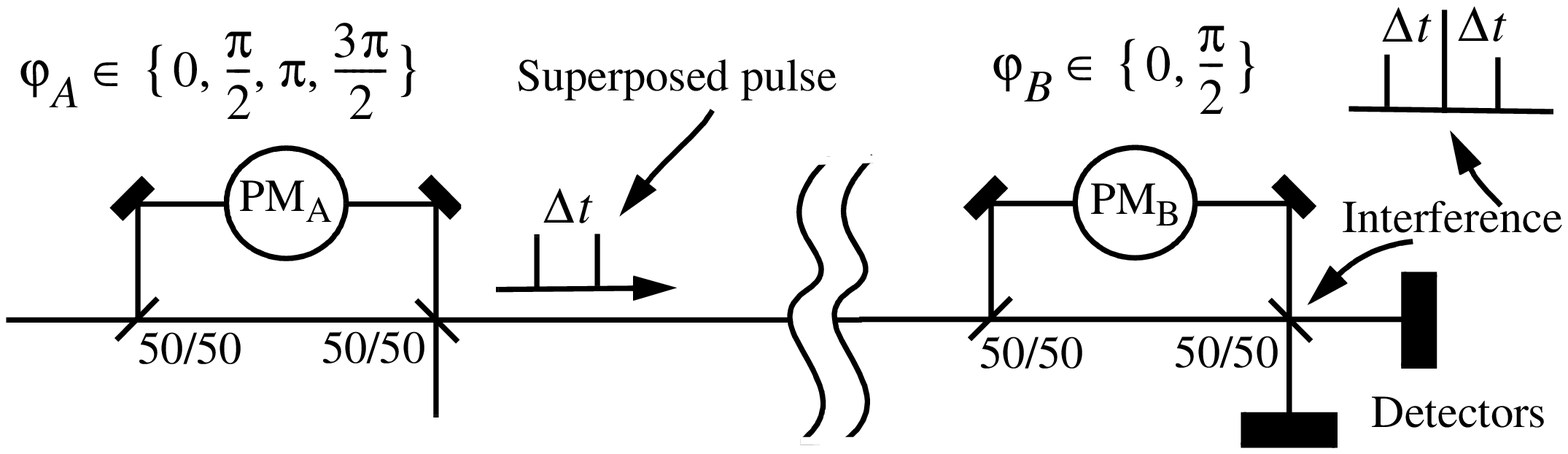,width=3.25 in}
\caption{Fiber based BB84-like transmitter and receiver. {\bf A} is on left,
{\bf B} on right. The phase modulator at {\bf A} is randomly toggled between
one of four  phases: $\phi_A \in \{0,\pi/2,\pi, 3 \pi/2 \}$, and at {\bf B} 
between one of  two phases: $\phi_B \in \{0, \pi/2 \}$.}
\label{fig:bb84}
\end{figure}

\begin{figure}[!h]
\psfig{figure=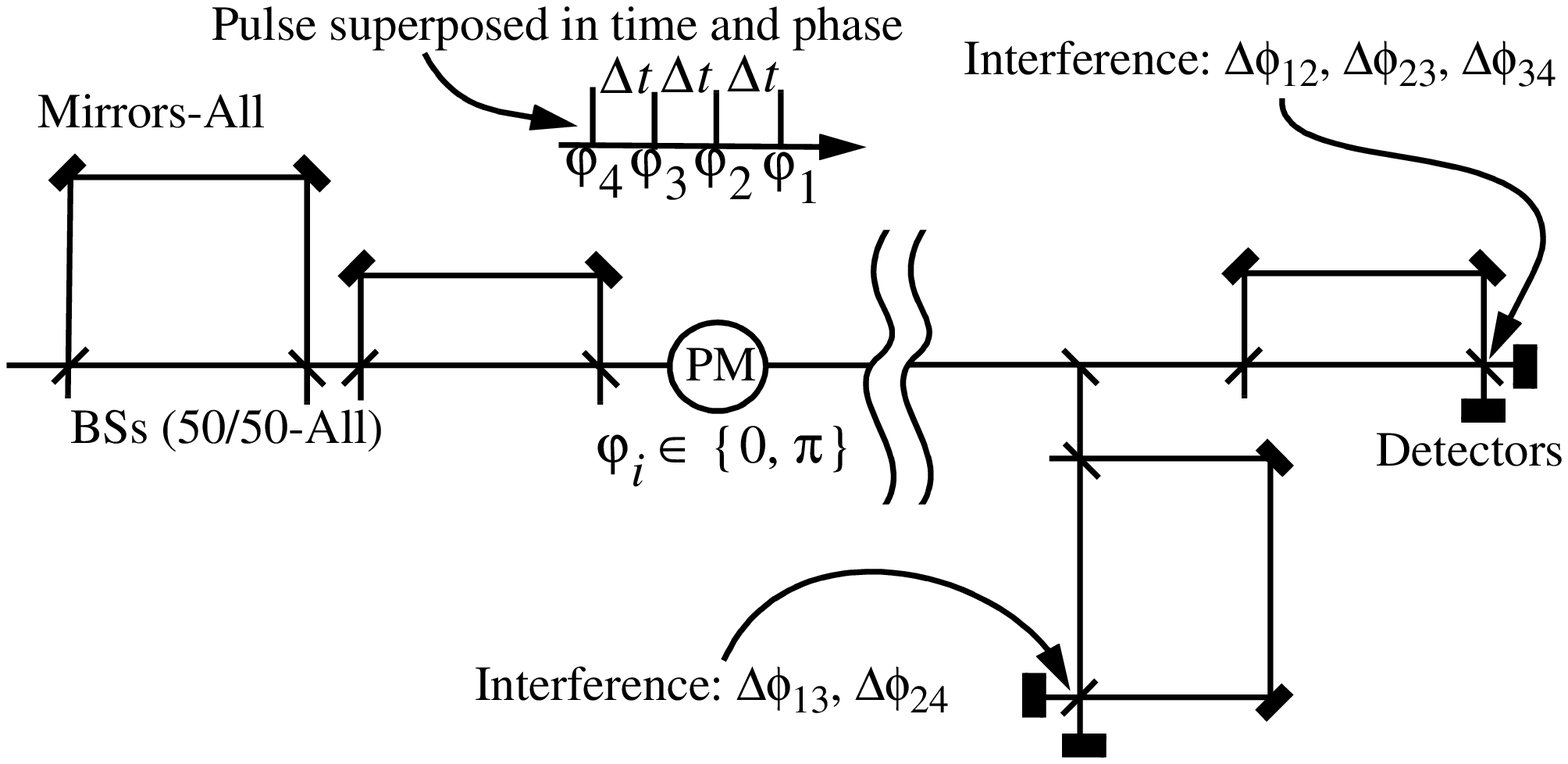,width=3.25 in}
\caption{A BLT series-delay interferometer variation with increased disturbance
and reduced potential eavedropper information. {\bf A} is on left, {\bf B} on 
right. This system increases the eavesdropping disturbance to as much as
$2/5$  and reduces {\bf E}'s information on the quantum key to a maximum of
$1/5$.}
\label{fig:blt01v3}
\end{figure}

\begin{figure}[!h]
\psfig{figure=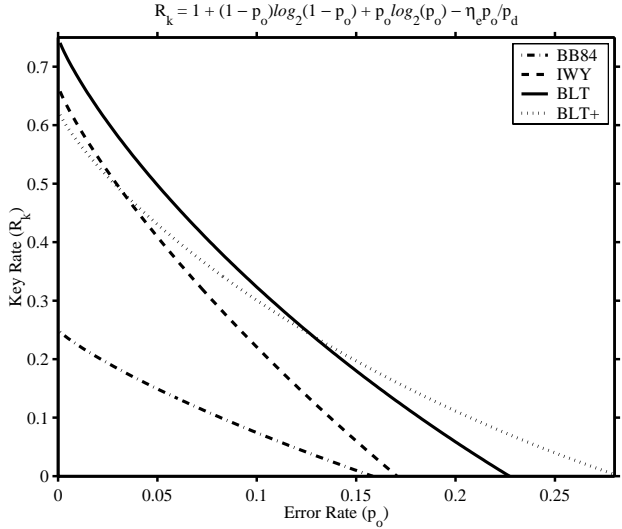,width=3.25 in}
\caption{Fraction of transmitted qubits which contribute to private key when 
privacy amplification is considered. This plot clearly shows that the  series 
element idea has a great advantage over BB84, IWY, or BLT$\oplus$ for error 
rates les than about $0.13$. (The values presented here assume true 
single-photon QKD and do not consider weak coherent attack strategies, or the 
amount of key lost through error reconciliation.)}
\label{fig:key-rate}
\end{figure}


\begin{thebibliography}{99}
\bibitem{ref:note4} The privacy amplification key-rate specified here relates 
how much key must be removed irrespective of the additonal amount of key lost 
during error reconciliation.
\end{thebibliography}
\end{document}